*Title page (with authors and addresses)

# Short term memory effects of an auditory biofeedback on isometric force control: Is there a differential effect as a function of transition trials?

Rémy CUISINIER*, Isabelle OLIVIER, Jocelyne TROCCAZ, Nicolas VUILLERME, Vincent NOUGIER

Laboratoire TIMC-IMAG, UMR UJF CNRS 5525, Grenoble, France.

* Address for correspondence:
Rémy CUISINIER
Laboratoire TIMC-IMAG, UMR UJF CNRS 5525
Faculté de Médecine
38706 La Tronche cedex, France.
Fax: +33 (0)4 76 63 51 00
Email: remy.cuisinier@imag.fr

Number of pages: 16
Number of figures: 3
Number of tables: 0

# Short term memory effects of an auditory biofeedback on isometric force control: Is there a differential effect as a function of transition trials?

**Abstract**

The aim of the present study was to investigate memory effects, force accuracy and variability during constant isometric force at different force levels, using an auditory biofeedback. Two types of transition trials were used: A Biofeedback-No biofeedback transition trial and a No biofeedback-Biofeedback transition trial. The auditory biofeedback produced a low- or high-pitched sound when participants produced an isometric force lower or higher than required, respectively. To achieve this goal, 16 participants were asked to produce and maintain two different isometric forces (30 ± 5% and 90 N ± 5%) during 25 s. Constant error and standard deviation of the isometric force were calculated.
While accuracy and variability of the isometric force varied according to the transition trial, a drift of the force appeared in the no biofeedback condition. This result suggested that the degradation of information about force output in the no biofeedback condition was provided by a leaky memory buffer which was mainly dependent on the sense of effort. Because this drift remained constant whatever the transition used, this memory buffer seemed to be independent of short term memory processes.

**Introduction**

It is well established that force production can be controlled by two different mechanisms. The first one is provided by the feedforward command of the central nervous system generally called sense of effort (McCloskey et al., 1974; Gandevia, 1996; Proske et al., 2004). The second one is provided by feedback signals from ascending sensory information and called sense of force (Simon and Ferris, 2008). Golgi tendons organs and cutaneous receptors can provide such information to gauge sense of force.

In some cases, the use of tools does not allow to provide good proprioceptive and haptic feedbacks. For example, currently available laparoscopic tool in minimally invasive surgery is unable to provide such information to control force production in robot-assisted surgical systems. Indeed, direct contact is replaced by long instruments between the tissue and fingers, resulting in reduced and distorted haptic (kinaesthetic and tactile) feedback (den Boer et al., 1999; Zahariev and Mackenzie, 2008). This can modify surgeon's behavior, especially in the application of forces while interacting with the tissue.

To supply the lack of sensory information in using tools, another type of feedback seemed to be necessary. In this way, sensory substitution has been used to replace direct tactile or force feedback with other sensory modalities such as visual or auditory feedback to provide a better representation of force production. In tasks requiring a steady force, it is possible to introduce an auditory biofeedback to control force production by providing information when the force is lower or higher than required. This auditory rather than visual biofeedback could be useful because in specific tasks like surgical ones when a computer aided surgery takes place, visual information is often used to control tools trajectories.

Several studies have shown that sensory substitution feedback enhances movement ability and control (Massimino, 1995; Kitawaga et al., 2005; Zahariev and Mackenzie, 2007, 2008). For example, Zahariev and Mackenzie (2007, 2008) showed that movement time shortened using an auditory cue to grasp an object with different sizes. Using a protocol with visual feedback withdrawal after several seconds (i.e., in vision-Novision transition trials), several studies found that variability of the isometric force production decreased with visual withdrawal when normalized by target force or maximal voluntary contraction (Tracy, 2007; Welsh et al., 2007) or increased without normalization (Davis, 2007; Baweja et al., 2009). These studies suggested that visual feedback contributes to increase force correction and thus force variability during an isometric force task. Moreover, these authors highlighted the presence of a drift of the isometric force over time which was dependent on force level after withdrawal of visual information. They excluded the effect of fatigue to explain this drift because they found it also for lower forces (5% MVC).

Among these authors, results have also shown that the initiation of force decay started 1.5 - 2.5 s following the removal of visual information (Vaillancourt et al., 2001; Vaillancourt and Russell, 2002; Davis, 2007). This time was too long to explain this force decay over time by the sensory reflex pathways. Indeed, studies have shown that changes in force output due to altered feedback conditions occur after 20-150 ms for proprioceptive feedback (Marsden et al., 1983). The decay over time better fit with the impact of short term memory processes. It could explain why the force decay occurred only after 1.5 – 2.5 s following the removal of visual inputs. Davis (2007) agreed with the impact of short memory processes on force output and extended this effect to a bimanual isometric finger force task. This author proposed a visuomotor force control model which includes a leaky memory buffer to explain the exponential drift of force output when no error feedback was available. In line with this finding, Vaillancourt et al. (2001) found that this decay of force over time was faster in parkinson's disease and concluded that this effect was related to deficits in higher sensory-motor processes.

The effect of the short term memory processes on motor control has been also described by Miall, Haggard and Cole (1995) by analyzing the position and amplitude of several wrist flexion and extension movements with and without vision as a function of pause interval between each trial. By plotting parameters of the first movement made after each pause against the pause interval, they obsverd a degradation of the accuracy related to the interval duration. Following a pause shorter than six seconds, the first movement was quite accurate. After 12 seconds, even the first movement was inaccurate in amplitude and in absolute position. Miall et al.'s result (1995) suggested that there is some form of internal visuo-motor memory supporting motor control and that the memory fades only after few seconds. In other words, the accuracy of reproduction of a required level of force without any biofeedback would be dependent on this memory effect after a short pause.

Therefore, the purpose of the present study was to investigate force accuracy and variability during constant isometric force production at different force levels in two types of transition trials. Biofeedback-Nobiofeedback transition trials were used to evaluate subjects' ability to regulate the isometric force while being exposed to abrupt changes in the availability of auditory information. Nobiofeedback-Biofeedback transition trials were used to evaluate subjects' ability to 1) reproduce the force required after a short period of time and 2) integrate a new auditory biofeedback to regulate their isometric force.

**Methods**

*Participants*

16 right-handed adults (age: 27 ± 6.1 years; body weight: 72.3 ± >8.2 kg; height: 178.1 ± 7.4 cm; mean ± SD) with no history of previous motor problems or neurological disease voluntarily participated in the experiment. They gave their informed consent to the experimental procedure as required by the Helsinki declaration (1964).

*Procedures*

The feet in a semi tandem position, participants stood in front of a force platform (AMTI®, model OR6-5-1) which was vertically positioned at the elbow level. The right arm flexed to 90 deg, the fist closed, they were asked to produce and maintain a horizontally isometric force on the force platform with or without an auditory feedback. This auditory feedback was provided by two continuous buzzers (low- and high-pitched sound, Velleman models, sound frequency of 2.8 kHz and 4.5 kHz, respectively).When participants produced a lower force than required, the low-pitched sound was delivered whereas a high-pitched sound occurred when participants produced a higher force than required. Participants were asked to produce two different forces: 30 N ± 5 % and 90 N ± 5 %. In these intervals (± 5 %), no auditory feedback was provided indicating that participants produced the required force.

Before the experiment, participants heard the auditory biofeedback to identify the low and high-pitched sounds. Ten 25 s trials for each level of force were then performed. Two types of trials which corresponded to two different feedback transitions were presented alternatively in each block of 10 trials per force level. In uneven trials, participants produced and maintained the force with the auditory feedback during the first 15 s. For the later 10 s, no auditory feedback was provided whatever the force produced (trials with a BioFB-NoBioFB transition) but participants were instructed to maintain the force required. In even trials, participants had to produce and maintain the required force during the first 15 s without auditory feedback. For the later 10 s, participants could gain efficiency by using the auditory feedback to reach the required force (trials with a NoBioFB-BioFB transition) (cf. figure 1). The rest time between two transition trials was 20 s. Participants reported no muscle fatigue during the protocol.

*Data analysis*

Force production was analyzed during two temporal frames of 8 s each: from 7 to 15 s and from 17 to 25 s called first and second temporal frames, respectively, in order to eliminate 1) the time necessary to produce and increase the force to reach the force target and 2) the time necessary to adjust the force in the NoBioFB-BioFB transition (i.e., from 15 to 17 s as observed in figure 1). Data forces were sampled at 100 Hz (12-bit A/D conversion). Data were low pass filtered with a second-order Butterworth (6 Hz). The cut-off frequency was fixed following a spectral and residual analysis (Winter, 1990).

Four dependent variables were used to describe participants' isometric force production behavior: (1) the percentage of success of the isometric force production (i.e., the percentage of time over the two temporal frames of 8 s spent in the required force target 30 N ± 5 % or 90 N ± 5 %), (2) the percentage of constant error (CE) expressed as the constant error (mean produced force minus force target) divided by force target, 3) the slope of the linear regression of the isometric force calculated for each individual trial and each temporal frame (7-15 s and 17-25 s) to quantify the drift of the isometric force production within a given time interval (Vaillancourt et al., 2002; Davis, 2007; Tracy, 2007). This drift was expressed in $N.s^{-1}$ and normalized with respect to the mean force produced. The fourth dependent variable was the coefficient of variation (SD/mean). For each force target and each temporal frame, this variable was quantified after removing the linear trend due to the drift of the isometric force control. The drift was removed because it did not represent the force fluctuations of interest and would increase SD values. This SD was then normalized as a percentage of the mean force produced during each force and each BioFB condition. Note that the normalization with respect to the force target was not applied due to the presence of the drift.

-------------------------
Insert Figure 1 about here
-------------------------

*Statistical analysis*

A previous analysis was done to test the existence of a learning effect. Five trials x 2 force targets (30 N ± 5 % vs. 90 N ± 5 %) × 2 temporal frames (7-15 s vs. 17-25 s) analyses of variance (ANOVAs) were applied to the two transition trials (NoBioFB-BioFB and BioFB-NoBioFB transitions). Results did not show any effect of trials on the variables used. As a result, the five trials were averaged for each experimental condition and used for further statistical analyses. Three ANOVAs with repeated measures were used to answer different questions.

In order to check that participants were as accurate in the first and second temporal frames when using the auditory Biofeedback (i.e., from 7 to 15 s in the BioFB-NoBioFB transition trials vs. 17 to 25 s in the NoBioFB-BioFB transition trials), 2 force targets (30 N ± 5 % vs. 90 N ± 5 %) × 2 temporal frames (7-15 s vs. 17-25 s) ANOVAs with repeated measures on all factors were applied to the data. The same analysis was done when the auditory biofeedback was not available in order to test the effect of a short pause to reproduce the required force. So, 2 force targets (30 N ± 5 % vs. 90 N ± 5 %) × 2 temporal frames (7-15 s in the NoBioFB-BioFB transition trials vs. 17-25 s in the BioFB-NoBioFB transition trials) ANOVAs with repeated measures on all factors were applied to the data.

In order to investigate the effects of adding the auditory Biofeedback, 2 force targets (30 N ± 5 % or 90 N ± 5 %) × 2 auditory feedbacks (NoBioFB *vs.* BioFB) ANOVAs with repeated measures on all factors were applied to the NoBioFB-BioFB transition trials (i.e., from 7 to 15 s vs. 17 to 25 s in the NoBioFB-BioFB transition trials). Finally, in order to investigate the effects of interrupting the auditory biofeedback on force control, 2 force targets (30 N ± 5 % or 90 N ± 5 %) × 2 auditory feedbacks (BioFB *vs.* NoBioFB) ANOVAs with repeated measures on all factors were applied to the BioFB-NoBioFB transition trials (i.e., from 7 to 15 s vs. 17 to 25 s in the BioFB-NoBioFB transition trials).

Post hoc analyses (*Newman-Keuls*) were performed whenever necessary. The level of significance was set at 0.05.

**Results**

*Force control when using the auditory biofeedback*

Analysis of the percentage of success did not show any main effect of force target or temporal frame ($F_{(1,15)} = 0.39$, $p > 0.5$ and $F_{(1,15)} = 2.90$, $p > 0.1$, respectively) and no interaction of the two factors ($F_{(1,15)} = 0.86$, $p > 0.3$). The percentage of success was about 70.95 % ± 1.76 (mean ± standard error) and was similar whatever the force target and temporal frame when using the auditory biofeedback.

Analysis of the percentage of constant error did not show any main effect of force target or temporal frame ($F_{(1,15)} = 4.16$, $p > 0.05$ and $F_{(1,15)} = 2.12$, $p > 0.1$, respectively) and no interaction of the two factors ($F (1,15) = 1.58$, $p > 0.2$). The percentage of constant error did not exceed 2.62 % ± 0.24.

Analysis of the coefficient of variation only showed a main effect of force target ($F (1,15) = 5.07$, $p < 0.05$). The coefficient of variation decreased from 30 N to 90 N (2.34 % ± 0.27 vs. 1.88 % ± 0.25, respectively).

Finally, the slope of the isometric force did not differ as a function of force target and temporal frame ($F (1,15) = 0.11$, $p > 0.7$ and $F (1,15) = 1.99$, $p > 0.15$, respectively). No interaction between the two factors was observed ($F (1,15) = 0.36$, $p > 0.5$).

*Force control in the absence of auditory biofeedback*

Analysis of the percentage of success showed a trend for the main effects of force target and temporal frame ($F (1,15) = 4.26$, $p = 0.057$ and $F (1,15) = 4.53$, $p = 0.05$, respectively) and a significant interaction of force target x temporal frame ($F (1,15) = 8.64$, $p < 0.05$). The decomposition of the interaction into its simple main effects showed that the percentage of success in 30 N was higher in the second (in the BioFB-NoBioFB transition trials) than in the first temporal frame (in the NoBioFB-BioFB transition trials). In other words, the percentage of success was higher when interrupting the auditory biofeedback than when initially producing the isometric force without biofeedback (44.23 % ± 5.15 *vs*. 23.26 % ± 4.35). Note that this percentage of success was also higher in the 30 N than in the 90 N force target condition whatever the temporal frame (44.23 % ± 5.15 *vs*. 28.79 % ± 2.84 and 25.49 % ± 5.52, respectively).

Analysis of the percentage of constant error showed a main effect of temporal frame ($F (1,15) = 10.50$, $p < 0.005$) and a significant interaction of temporal frame x force production ($F (1,15) = 6.03$, $p < 0.05$). The decomposition of this interaction into its simple main effects showed that the percentage of constant error did not vary in the 90 N force target condition (8.48 % ± 0.82 *vs*. 8.04 % ± 0.95) whereas this percentage was smaller in the second (in the BioFB-NoBioFB transition trials) than in the first temporal frame (in the NoBioFB-BioFB transition trials) in the 30 N force target condition (14.46 % ± 2.91 *vs*. 5.93 % ± 1.20, respectively). In other word, the percentage of constant error was smaller when the 30 N isometric force was maintained after the interruption of the auditory biofeedback than when initially producing the 30 N force without biofeedback.

Analysis of the coefficient of variation only showed a main effect of temporal frame ($F (1,15) = 6.60$, $p < 0.05$). This coefficient of variation was smaller when maintaining the force after the interruption of the auditory biofeedback than when starting to produce the required isometric force without biofeedback (2.09 % ± 0.19 *vs*. 1.67 % ± 0.20, respectively).

Finally, analysis of the slope of the isometric force did not show any main effect of force target ($F (1,15) = 1.42$, $p > 0.25$), temporal frame ($F (1,15) = 0.38$, $p > 0.5$) or interaction of the two factors ($F (1,15) = 0.00$, $p > 0.90$).

*Force control when introducing the auditory biofeedback (NoBioFB-BioFB transition trials)*

Analysis of the percentage of success only showed a main effect of biofeedback ($F (1,15) = 571.76$, $p < 0.001$). As observed in figure 2, participants were more successful with than without the auditory biofeedback whatever the force required (70.05 % ± 1.97 vs. 26.02 % ± 3.68, respectively).

Analysis of the percentage of constant error showed a main effect of biofeedback ($F (1,15) = 47.18$, $p < 0.001$) and a significant interaction of force target x biofeedback ($F (1,15) = 4.62$, $p < 0.05$). The decomposition of the interaction into its simple main effects showed that with the auditory biofeedback the percentage of constant error was similar whatever the force target (2.34 ± 0.25 % vs. 3.09 ± 0.27 % for 30 N and 90 N, respectively). Conversely, without biofeedback the percentage of constant error was higher for the 30 N than for the 90 N force target suggesting that force control was more difficult for lower levels of force (14.46 ± 2.91 % vs. 8.48 ± 0.83 %, respectively - figure 2).

The same analysis applied to the coefficient of variation did not showed any effect of the force target ($F (1,15) = 0.59$, $p > 0.45$), biofeedback ($F(1,15) = 0.30$, $p > 0.50$), and interaction of the two factors ($F(1,15) = 0.95$, $p > 0.30$). Note that the coefficient of variation was about 2.16 ± 0.25 %.

Analysis of the slope of the linear regression of the isometric force only revealed a significant main effect of biofeedback ($F (1,15) = 17.87$, $p < 0.001$). The decay of force over time disappeared when the auditory biofeedback was available, i.e., normalized from -0.56 ± 0.17 to .11 ± 0.05, respectively (figure 3B).

--------------------------
Insert Figure 2 about here
--------------------------

*Force control when interrupting the auditory biofeedback (BioFB-NoBioFB transition trials)*

Analysis of the percentage of success showed main effects of force target ($F (1,15) = 14.80$, $p < 0.005$) and biofeedback ($F (1,15) = 79.87$, $p < 0.001$) and an interaction of force target x biofeedback ($F$ (1,15) = 12.21, $p < 0.005$). The decomposition of the interaction into its simple main effects showed a decrease of the percentage of success with the interruption of the auditory biofeedback. This decrease was larger for the 90 N than for the 30 N force target (from 71.40 ± 1.49 % to 25.49 ± 5.52 % *vs.* from 71.77 ± 1.58 % to 44.23 ± 5.14 %, respectively, p<.005) (figure 2).

Analysis of the percentage of constant error only showed a main effect of biofeedback ($F$ (1,15) = 34.61, $p < 0.001$). As illustrated in figure 2, this percentage increased when interrupting the auditory biofeedback (2.51 ± 0.21 *vs.* 6.99 ± 1.09 %, respectively).

Analysis of the coefficient of variation showed main effects of force target ($F$ (1,15) = 9.34, $p < 0.01$) and biofeedback ($F$ (1,15) = 9.42, $p = 0.01$). A decreasing variability was observed from 30 N to 90 N (2.07 % ± 0.22 vs. 1.60 % ± 0.15, respectively) and when interrupting the auditory biofeedback (2.00 % ± 0.21 vs. 1.67 % ± 0.16, respectively).

Analysis of the slope of the linear regression of the isometric force only revealed a significant main effect of biofeedback ($F$ (1,15) = 18.40, $p < 0.001$). The decay of force over time appeared after interrupting the auditory biofeedback (normalized slope from -0.47 ± 0.15 to 0.02 ± 0.06, respectively) (figure 3B).

--------------------------
Insert Figure 3 about here
--------------------------

**Discussion**

*Force accuracy, force decay and biofeedback transitions*

The purpose of the present study was to investigate the transitional effects when using or interrupting an auditory biofeedback to control an isometric force production. To achieve this goal, two alternative trials were presented: Trials with a biofeedback-no biofeedback transition (BioFB-NoBioFB) and trials with a no biofeedback-biofeedback transition (NoBioFB-BioFB). The underlying principle of this auditory feedback was to deliver low- and high-pitched sounds when force production was lower or higher than required, respectively. In addition, two low and high levels of force (30 N and 90 N) were manipulated in order to investigate to which extent the effects of the biofeedback were dependent on the level of force production. Reinserting auditory information could lead participants to integrate auditory input with proprioceptive inputs, whereas the withdrawal of auditory biofeedback forced participants to reorganize sensory information because proprioceptive information became the only remaining source of information for regulating the isometric force. We hypothesized differences in regulating isometric force as a function of the transition trial.

Results showed that participants spent more time in the force target, i.e., better maintained the required force over time and reduced the constant error when using the auditory biofeedback than when it was absent, whatever the force required (30N ± 5% or 90N ± 5%). This was not surprising and corroborated previous results on the effects of a visual biofeedback to improve force control (Vaillancourt and Russell, 2002; Gerovich et al., 2004; Wagner et al., 2007; Reiley et al., 2008). The present result extended this improvement to the use of an auditory biofeedback. It suggested that participants were able to integrate artificial force production information delivered through a two tones auditory biofeedback to produce and maintain a required level of force whatever the transition used.

When comparing the no biofeedback condition in the NoBioFB-BioFB versus BioFB-NoBioFB transition trials for the 30 N force, a lower percentage of success was observed because of a larger percentage of constant error. It seems that participants were less accurate to reproduce a lower 30 N force without any biofeedback as compared to a higher 90 N force production. This effect was also reported by Gandevia and Kilbreath (1990) in a force matching task (3 vs. 15 % of maximal isometric force for each muscle group) using proximal and distal muscles of the upper limb (first dorsal interosseous, flexor longus pollicis, biceps brachii). They showed a better accuracy using proximal than distal muscles and heavy than light weights. They excluded the impact of reflex inputs from muscles, joint and cutaneous receptors because the distal extremities (i.e., fingers) having the highest tactile acuity were not the most accurate for the matching weight task.

In summary, our results confirmed those reported by Gandevia and Kilbreath (1990) suggesting that 1) the recruitment of many motor units is not necessarily associated to an increased accuracy of force estimation and 2) with a progressive motor units recruitment as for a low force, the ability to grade muscular force does not remain constant. This unexpected findings need to be clarified with further experimentations.

Furthermore, in conditions in which auditory biofeedback was withdrawn, output force steadily declined. Our results did not differ from previous studies in which this decay over time was compensated for by the visual biofeedback whatever the transition used. However, we found that a linear fit better fit with this decay over time than an exponential fit. This effect was probably due to the methodology presently used. We recorded only ten seconds of force production without feedback while the temporal frame used in previous studies was larger (Vaillancourt and Russell, 2002; Davis 2007). It is clear however, that time force production cannot decline below a given level of force production. In other words, our temporal frame was not large enough to report the exponential decay of force production.

To explain this drift, Vaillancourt and Russell (2002) proposed that visuomotor force control includes a leaky memory buffer. Using another sensory input to provide the biofeedback, the present results extended the integration of this leaky memory buffer to another sensory motor control and confirmed that this buffer is a central resource (Davis, 2007). Moreover, when normalizing the slope decay by the force output, we observed that this decay did not differ as a function of force output and temporal frame (8-15 s or 17-25 s). Put together, these results suggested that the degradation of information about force output provided by this leaky memory buffer was mainly dependent on the sense of effort. This finding was in line with previous results showing that the decay of force over time was more pronounced in deafferented patients (Lafargue et al., 2003) and parkinson's disease (Vaillancourt et al., 2001) than in control subjects. However, because this drift remained constant whatever the transition used, this memory buffer seemed to be rather independent of short term memory processes. More generally, the key questions probably regard (1) and (2) the delay of efficiency of the short-term memory processes.

*Force variability and biofeedback transitions*

We observed the same effect of force fluctuation with an audio- as with a visuo-motor correction of the isometric force task (Vaillancourt et al., 2001; Vaillancourt and Russell, 2002; Davis, 2007). This was true even though visual afferent information was available during the entire force production task in previous studies, allowing continuous adjustments, whereas in the present study, auditory afferent information was provided only when the produced force was out of the force target. This general result may be due to the target used (± 5%) which was not large enough to reduce the variability, i.e., the number of adjustments required to remain in the target.

For example, several authors have found a non linear relationship between the gain of the visual feedback and force variability. More precisely, it has been shown that an inverted U shaped curve took place between these two parameters. Increasing the gain up to a given threshold enhanced the motor performance but increasing the gain beyond this threshold could lead to performance deficits (Sosnoff et al., 2006; Hong and Newell, 2008). However, this hypothesis is contradictory with recent results suggesting the absence of a U-shape function between gain and force (Baweja et al., 2009; Prodoehl and Vaillancourt, 2010). Four factors, at least, could explain the discrepancy observed in the literature: (1) The type of feedback used (visual or auditory), (2) the intensity of the required force output which varied across experiments, (3) the type of movement required for producing the force output, and (4) the presence or absence of a bandwidth target which may weaken the gain effects and the U-shaped finding.

Previous studies comparing force variability with and without visual feedback during constant isometric tasks exhibited some contradictory results. Studies which involved the use of fingers, and thus primarily small hand and forearm muscles, suggested that the removal of visual feedback does not influence force variability (Vaillancourt and Russell, 2002). Other studies which involved primarily larger muscles (Dorsiflexion or plantarfexion of the foot, quadriceps femoris) suggested that the removal of visual feedback decreases force variability (Tracy, 2007; Welsh et al., 2007). However, using index finger abduction, Baweja et al. (2009) found an increasing variability with the use of the visual biofeedback and so, rejected the effect of muscles size to explain this difference.

In the present study, the presence of the auditory biofeedback was manipulated in two types of transition trials. In the BioFB-NoBioFB transition trials, we observed a decreasing variability of force output with the removal of the auditory biofeedback (from 2.07 % to 1.60 %), whereas in the NoBioFB-BioFB transition trials, the variability did not differ with the appearance of the biofeedback (about 2.16 %). In other words, the variability of the motor command only decreased with the removal of the auditory biofeedback. Two hypotheses could explain this result.

Following the first hypothesis, the decreasing variability illustrated the removal effect of the auditory biofeedback. Removing this feedback forced the subjects to reorganize sensory information because proprioceptive information became the only remaining source of information for regulating the isometric force. The decreasing variability observed in the BioFB-NoBioFB transition trials could result from the decreasing number of sensory sources of information to control the isometric force, each sensory input producing some variability. Comparison of this variability to the one observed in the NoBioFB-BioFB transition trials forced to reject this hypothesis. Indeed, if this hypothesis was true, the same variability would be observed when the task began by the no biofeedback condition contrary to what was observed.

The second hypothesis thus suggested that this effect was task-specific: Starting with the biofeedback immediately gave a target reference. To be successful, subjects could adjust their motor command only by decreasing the variability of force output. In other words, the withdrawal of the auditory biofeedback forced the subjects to adapt their motor planning with the only remaining source of information for regulating the isometric force. In the NoBioFB-BioFB transition trials, the task was first to reproduce the isometric force output based on the proprioceptive information, only. Inserting the auditory feedback could lead subjects to integrate auditory inputs with proprioceptive ones in order to adjust the gain necessary to reach the target force. Further experiment would be necessary to validate this explanation.

To conclude, isometric force control differed as a function of biofeedback transition. Interrupting a biofeedback forced the subjects to adapt their motor control with the only remaining source of sensory inputs whereas reproducing the required force and integrating a biofeedback during the force production only modified the gain of force. It is interesting to note that the decay over time did not differ as a function of the temporal frame used suggesting that the effect of short term memory processes was independent of those implied to reproduce a required force.

**Acknowledgements**

This project was supported by a public grant from the French research agency, through the TELEOS program (ANR-06-BLAN-0243).

**Legends**

Figure 1. Illustration for one subject of the average of five NoBioFB-BioFB transition trials (i.e., NoBioFB with the first temporal frame from 7 to 15 s and BioFB with the second temporal frame from 17 to 25 s) for a 90 N force target. In the BioFB-NoBioFB transition trials, BioFB was presented during the first 15 s and NoBioFB from 15 to 25 s.

Figure 2. Mean and standard error of A) the percentage of success and B) the constant error of the force production produced in the two temporal frames of analysis (7-15 s and 17-25 s) without and with auditory feedback (NoBioFB and BioFB, respectively), as a function of force target 30 N (*white bars*) and 90 N (*black bars*).

Figure 3. A) Example for one subject of the average of five NoBioFB-BioFB transition trials for the 90 N force target in which a drift of force over time was present in the NoBioFB condition and B) Mean and standard deviation of the slope of the linear regression of the isometric force in the NoBioFB-BioFB transition trials and BioFB-NoBioFB transition trials. The 30 N force target is represented with white bars whereas the 90 N force target is in black bars.

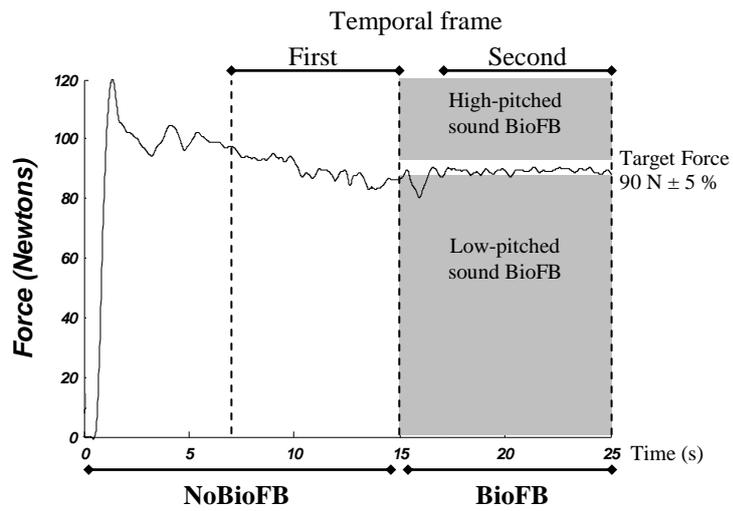

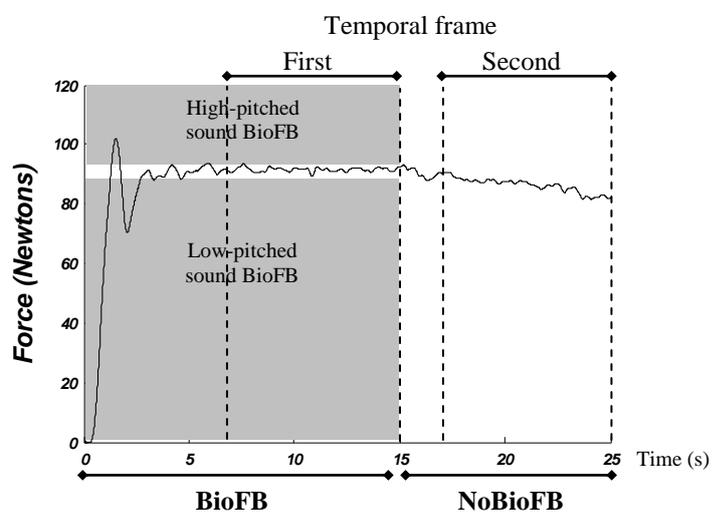

Figure 1

Figure 2

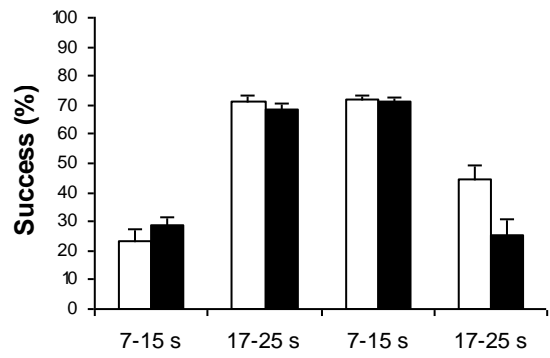

Figure 3

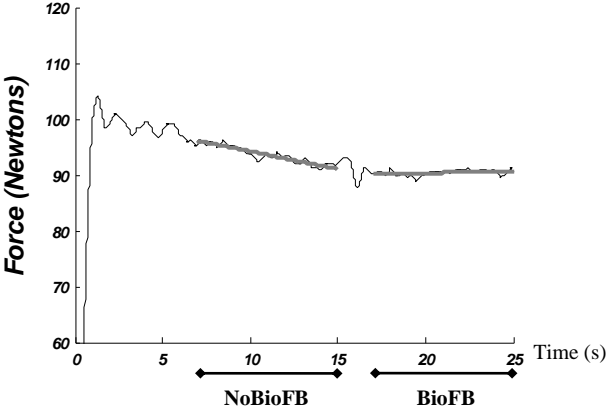